\documentclass[sigconf]{acmart}

\makeatletter
\def\@ACM@checkaffil{
    \if@ACM@instpresent\else
    \ClassWarningNoLine{\@classname}{No institution present for an affiliation}%
    \fi
    \if@ACM@citypresent\else
    \ClassWarningNoLine{\@classname}{No city present for an affiliation}%
    \fi
    \if@ACM@countrypresent\else
        \ClassWarningNoLine{\@classname}{No country present for an affiliation}%
    \fi
}
\makeatother

\usepackage{booktabs} 
\usepackage{multirow}
\usepackage{color}
\usepackage{array}
\usepackage{breqn}
\usepackage{{inputenc}}

\usepackage{balance}
\usepackage{multirow,tabularx}
\usepackage{longtable}
\usepackage{booktabs,longtable}
\usepackage{hhline}
\usepackage[flushleft]{threeparttable}
\usepackage[ruled, lined, linesnumbered, commentsnumbered, longend]{algorithm2e}
\usepackage{epsfig}
\usepackage{graphicx}
\usepackage{epstopdf}
\usepackage{thmtools}
\usepackage{enumitem}
\usepackage{verbatim}
\usepackage{amsfonts}
\usepackage{hyperref}
\usepackage{arydshln}
\usepackage{caption}
\usepackage{subcaption}
\usepackage{svg}
\usepackage{amsmath}
\usepackage{enumitem}
\hypersetup{colorlinks=false,linkcolor=blue,urlcolor=blue,citecolor=red}
\usepackage{subcaption}
\usepackage{cleveref}
\newcommand{\tocite}[1]{\textcolor{red}{\textbf{CITE}}}
\AtBeginDocument{%
  \providecommand\BibTeX{{%
    \normalfont B\kern-0.5em{\scshape i\kern-0.25em b}\kern-0.8em\TeX}}}

\setcopyright{acmcopyright}
\copyrightyear{2023}
\acmYear{2023}
\acmDOI{XXXXXXX.XXXXXXX}

\acmConference{Under review.}
\acmPrice{}
\acmISBN{978-1-4503-XXXX-X/18/06}

\begin{document}

\title{Model-enhanced Contrastive Reinforcement Learning for Sequential Recommendation}

\author{Chengpeng Li}
\affiliation{%
  \institution{University of Science and Technology of China}}
\email{chengpengli@mail.ustc.edu.cn}

\author{Zhengyi Yang}
\affiliation{%
  \institution{University of Science and Technology of China}}
\email{yangzhy@mail.ustc.edu.cn}

\author{Jizhi Zhang}
\affiliation{%
  \institution{University of Science and Technology of China}}
\email{cdzhangjizhi@mail.ustc.edu.cn}

\author{Jiancan Wu}
\affiliation{%
  \institution{University of Science and Technology of China}}
\email{wujcan@gmail.com}

\author{Dingxian Wang}
\affiliation{%
  \institution{University of Technology Sydney}}
\email{dingxianwang@etsy.com}

\author{Xiangnan He}
\affiliation{%
  \institution{University of Science and Technology of China}}
\email{xiangnanhe@gmail.com}

\author{Xiang Wang}
\affiliation{%
  \institution{University of Science and Technology of China}}
\email{xiangwang1223@gmail.com}
\authornote{Corresponding author}

\renewcommand{\shortauthors}{Chengpeng Li et al.}

\begin{abstract}
In recent years, reinforcement learning (RL) has been widely applied in recommendation systems due to its potential in optimizing the long-term engagement of users.
From the perspective of RL, recommendation can be formulated as a Markov decision process (MDP), where recommendation system (agent) can interact with users (environment) and acquire feedback (reward signals).
However, it is impractical to conduct online interactions with the concern on user experience and implementation complexity,
 and we can only train RL recommenders with offline datasets containing limited reward signals and state transitions. 
Therefore, the data sparsity issue of reward signals and state transitions is very severe, while it has long been overlooked by existing RL recommenders.
Worse still, RL methods learn through the trial-and-error mode, but negative feedback cannot be obtained in implicit feedback recommendation tasks, which  aggravates the overestimation problem of offline RL recommender.

To address these challenges, we propose a novel RL recommender named model-enhanced contrastive reinforcement learning (MCRL). 
On the one hand, we learn a value function to estimate the long-term engagement of users, together with a conservative value learning mechanism to alleviate the overestimation problem.
On the other hand, we construct some positive and negative state-action pairs to model the reward function and state transition function with contrastive learning  to exploit the internal structure information of MDP.  
Specifically, we learn two predictive networks where positive state-action pair can lead the reward and next state in the dataset, and negative state-action pairs fail. 
Finally, regarding the reward and transition model learning as an auxiliary tasks, we use value-weighted regression to generate the recommendation policy.
In this way, we alleviate the problem of data sparsity and overestimation. 
Experiments demonstrate that the proposed method significantly outperforms existing offline RL and self-supervised RL methods with different representative backbone networks on two real-world datasets.

\end{abstract}





\maketitle

\section{Introduction}




Sequential recommendation (SeqRec) aims to predict the next item a user will interact with, based on her/his historical interaction sequences \cite{YoutubeRS, DIN, music, CL4Rec}.
Traditional sequential recommenders are usually trained by an auto-regressive paradigm, that is, to recover historical interaction sequence by optimizing the maximum likelihood probability of the next item step by step \cite{Caser,GRU4Rec,SASRec}.
This greedy optimization approach targets optimizing immediate engagement such as click-through \cite{GRU4Rec} rates, while ignoring the long-term engagement of users \cite{Resact}.
Realizing this issue, many researchers draw inspiration from reinforcement learning (RL) \cite{RL1,RL2,RL3,RL4}.
With long-term cumulative rewards as the goal of optimization, RL is more in line with the demands of a real recommendation system that pursues long-term interests.

\begin{figure}[htbp]
    \centering
    \includegraphics[width=1\linewidth]{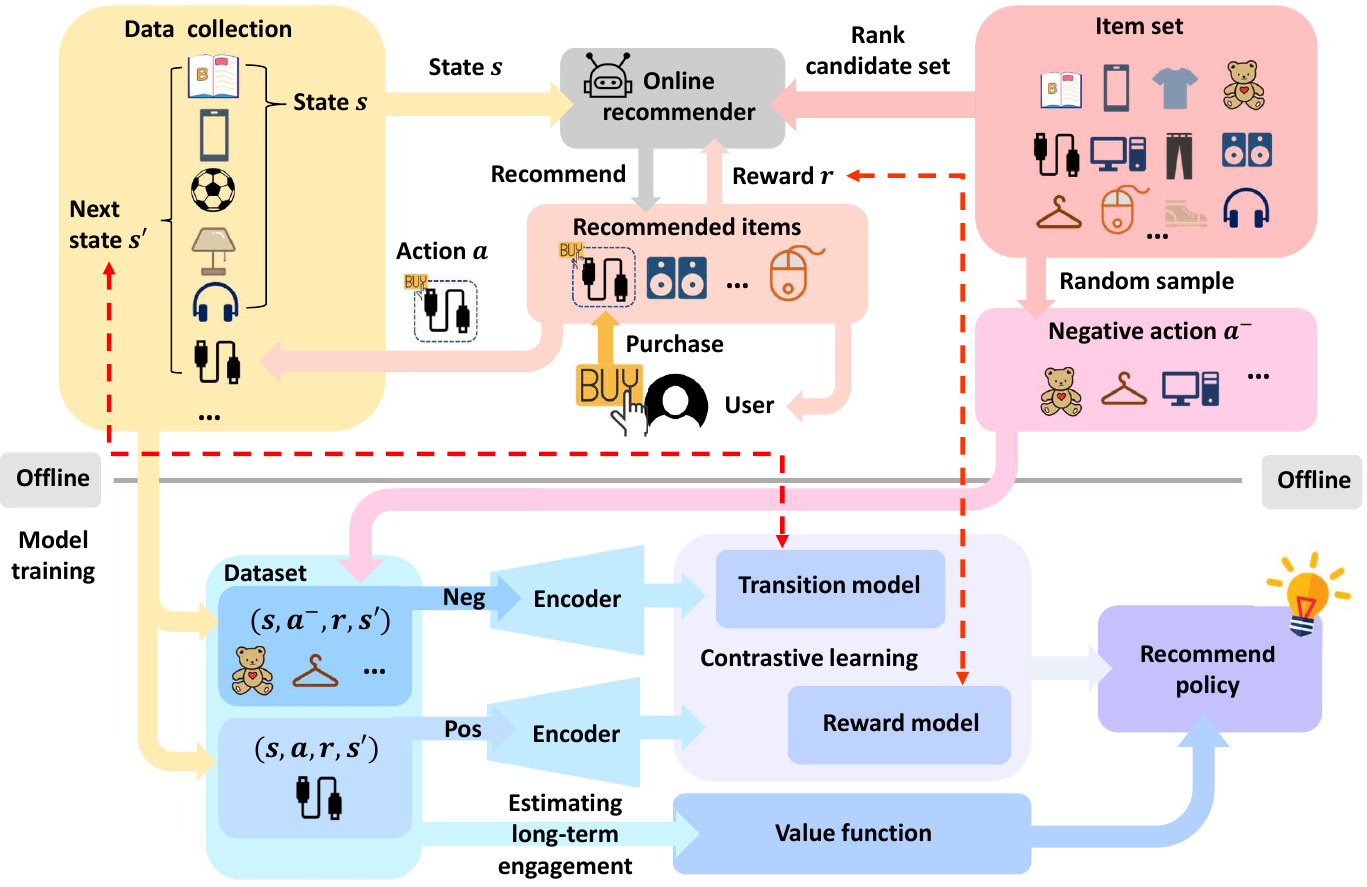}
    \caption{The whole process of MCRL. Above the gray line is the Data collection process during the interaction process between recommender (agent) and  user (environment). Below the gray line is the training procedure of MCRL. The red dotted line represents the transition model and reward model and corresponding MDP component.}
    \label{MDP}
\end{figure}

Taking the users as the environment, the recommender as the agent, and the real-time feedback of users as the reward signal, we can frame SeqRec as a Markov decision process (MDP), which is the foundation of RL \cite{MDPRS}. Figure \ref{MDP} illustrates the procedure.
However, conventional online RL approaches adopt the trial-and-error strategies to learn through constant interaction with the environment \cite{RL1,RL3}, which is impractical in recommendation since online interaction is too expensive \cite{offlinesurvey,Resact}. 
In addition, the cold start training may also harm the user experience.  
Therefore, offline RL is more suitable for recommendation systems.
Generally, studies on offline RL propose to restrict the agent from taking risky actions \cite{BCQ,BEAR,IQL,xin}, or constrain the estimation of value functions \cite{CQL,offlinesurvey} to make the recommender more conservative, and alleviate the overestimation of value functions in recommendation \cite{BCQ,BEAR}. 
However, current RL recommenders hardly consider the data sparsity of reward signal and state transition, which are important for RL methods, since it usually suffers from high variance and needs as much data as possible for accurate estimation of long-term engagement \cite{MBRL,variance}. 

Here we focus on offline RL for SeqRec.
By scrutinizing this research line, we find several inherent limitations:
\begin{itemize}[leftmargin=*]
    \item Sparsity of reward and state transition. As the recommender cannot interact with users to collect new data in real time, it can only utilize the observed interactions for policy learning. However, compared the huge state and action space, the observed interactions are extremely sparse in SeqRec, thus making it difficult for RL to learn a qualified representation of the state\cite{MBRL}. 
    \item Overestimation of value function. In the offline RL setting, the value function is usually optimized on a small part of the action space, but evaluated on all valid actions. This easily results in the inaccuracy of value function estimation \cite{BCQ,offlinesurvey}. Although the bias is not always positive, in combination with the maximization operator in RL, there can be a persistent overestimation of value function\cite{overestimation} and harm the final performance. 
    \item Ignorance of negative signal. In addition to positive samples, negative samples in recommendation can also provide rich collaborative signals to help the RL model learning by trial and error \cite{xin2}. However, in most existing RL-based studies \cite{xin,IQL,BCQ}, negative samples are not fully utilized, which is easy to cause the model collapsing problem --- that is, the model can easily be optimized by predicting all interactions into positive ones. 
\end{itemize}

To tackle these challenges caused by the data sparsity problem, we believe that it is possible to develop an RL-based SeqRec model to exploit the internal structure information of MDP (\textit{i.e.,} reward function and state transition function). 
Specifically, we propose a new training framework, MCRL (\underline{M}odel-enhanced \underline{C}ontrastive \underline{R}einforcement \underline{L}earning), to optimize the long-term engagement in SeqRec, which can be easily integrated into existing SeqRec models. 
There are two special designs for addressing the challenges. 
Firstly, inspired by the development of visual RL \cite{VRL1,VRL2,VRL4}, we model reward function and state transition (\textit{i.e.,} learning a network to predict next state and reward with current state-action pairs) and regard them as auxiliary tasks to boost the state representation in the huge discrete state space. By this way, more self-supervised signal from the internal structure information of MDP can be utilized to mitigate data sparsity. Secondly, we take items that are not interacted with by users as negative actions to construct positive and negative state-action pairs for model learning.  Benefiting from the negative signal, the overestimation issue can be alleviated implicitly. It is worth noting that our approach cannot be attributed to moldel-based RL methods, as we do not utilize the learned model to generate virtual state transitions and reward signals, but rather view it as an auxiliary task. Then we design a simple and effective value function learning to estimate the long-term engagement of different states. Finally, using the value-weight regression \cite{IQL}, we combine the value function estimation and imitation learning to extract the policy with the auxiliary tasks of learning reward model and state transition model. 





The main contributions of this work are summarized as follows:
\begin{itemize}[leftmargin=*]
    \item  We propose a new RL recommender named MCRL, which  
    utilize the structure information of MDP to mitigate the data sparsity issues.
    \item We introduce a new contrastive learning paradigm in MCRL for reward and state transition modeling from the perspective of negative action.
    \item We conduct extensive experiments to demonstrate that the proposed MCRL can significantly outperform existing offline RL and self-supervised RL recommenders with different representative backbone networks.   
\end{itemize}




\section{Task Formulation}
We first introduce the concept of sequential recommendation (SeqRec) and formulate the task as a Markov Decision Process (MDP) as well as a corresponding reinforcement learning (RL) setting. 

In a SeqRec scenario, we typically have historical sequential or session-based user data. Let $\mathcal{I}$ denote the item set. A user-item interaction sequence can be denoted as ${x}_{1:t} = \{x_1, x_2, \ldots, x_t\}$, where $x_i\in \mathcal{I},(0<i\leq t)$ is the $i_{th}$ item interacted with by the user.
When a user starts a session, the recommendation system feeds items to the user one by one according to the user's feedback(\textit{e.g.}, clicks and purchases in an e-commerce scenario or viewing time in a social media scenario). The recommendation system aims to recommend the most relevant item $x_{t+1}$ based on the user's historical interaction ${x}_{1:t}$. Moreover, an ideal recommendation system would be expected to improve the long-term engagement of users. The SeqRec process can be formulated as a Markov Decision Process (MDP), defined by a tuple $\langle \mathcal{S},\mathcal{A},\mathcal{P},\mathcal{R},\gamma\rangle$:
\begin{itemize}[leftmargin=*]
    \item State space $\mathcal{S}:$ A state $s_t\in \mathcal{S}(t>0)$ is defined as  a  user-item interaction sequence before timestamp $t$, \textit{i.e.},$s_t = x_{1:t}$. We can obtain the representation of the state by a sequential model $G(\cdot)$, \textit{i.e.}, $z_t = G(s_t)$. 
    \item Action space $\mathcal{A}:$ An action $a_t\in A$ represents the recommended
item. From offline data, we can extract $a_t$ at timestamp  $t+1$, \textit{i.e.} $a_t:=x_{t+1}$.
    \item State transition function $\mathcal{P}:\mathcal{S}\times\mathcal{S}\times\mathcal{A}\rightarrow \mathbb{R}$:   the state transition  probability $p(s_{t+1}\mid s_t,a_t)$ represents the probability density of transition to the next
state $s_{t+1}\in \mathcal{S}$ from the current state $s_t\in\mathcal{S}$ recommending an item $a_t$.  
    \item Reward function $\mathcal{R}:\mathcal{P}:\mathcal{S}\times\mathcal{A}\rightarrow \mathbb{R}$: the reward $r(s_t,a_t)$ is the  immediate reward after recommending an item $a_t$ to user at state $s_t$.  
    \item Discount factor $\gamma:$ the discount factor $\gamma\in [0,1]$ determines the present value of future rewards. If $\gamma = 0$, the recommendation system would be concerned only with maximizing immediate rewards. If $\gamma=1$, the recommendation system gives equal weight to future rewards as it does to immediate rewards.
\end{itemize}
In RL, the  value of taking action $a_t$ in the state 
$s_t$  under a policy $\pi(a_t\mid s_t):\mathcal{S}\times\mathcal{A}\rightarrow  [0,1]$ , denoted by $Q_{\pi}(s,a)$, is defined as the expected cumulative reward (return) starting from state $s_t$, taking the action $a_t$, and then following the policy $\pi$:
\begin{equation}
\label{Qfunction}
Q_\pi\left(s_t, a_t\right)=\mathbb{E}_{\left(s_{t^{\prime}}, a_{t^{\prime}}\right) \sim \pi}\left[r\left(s_t, a_t\right)+\sum_{t^{\prime}=t+1}^{\infty} \gamma^{\left(t^{\prime}-t\right)} \cdot r\left(s_{t^{\prime}}, a_{t^{\prime}}\right)\right].
\end{equation}
Similarly, we define the value of a state $s_t$ under a policy $\pi$, denoted by $V_{\pi}(s_t)$,  as the expected return starting in $s_t$ and thereafter following policy $\pi$:
\begin{equation}
\label{vfunction}
V_\pi\left(s_t\right)=\mathbb{E}_{\left(s_{t^{\prime}}, a_{t^{\prime}}\right) \sim \pi}\left[\sum_{t^{\prime}=t}^{\infty} \gamma^{\left(t^{\prime}-t\right)} \cdot r\left(s_{t^{\prime}}, a_{t^{\prime}}\right)\right].
\end{equation}
The optimization objective is to find the optimal policy $\pi^*$ to maximize the expected return:
\begin{equation}
\pi^* = \mathop{\arg\max} _\pi \mathbb{E}_{(s_t,a_t) \sim \rho_{\pi}}\left[V^\pi\left(s_t\right)\right],
\end{equation}
where $ \rho_{\pi}$  denotes the state marginals of the trajectory distribution induced by a policy $\pi$.

\section{METHODOLOGY}

In this section, we describe our proposed MCRL, in detail. 
First, we introduce a value function estimator of the long-term value of various states in the recommendation system. 
We use a conservative TD loss to ease the overestimation problem. Next, we extract the policy based on the learned value function. To solve the problem of sparse data and lack of negative feedback in RL recommender, we reconstruct the reward function and state transition through contrastive learning during policy extraction, to enhance the 
state representation. 
Figure \ref{MDP} illustrates the whole process of MCRL.

\subsection{Long-term Engagement Estimation via Value Function}

Since we have modeled the SeqRec task as RL, the estimation of long-term benefits is correspondingly equivalent to the learning of the value function $V_{\pi}(s_t)$ and Q function $Q_{\pi}(s_t,a_t)$. As the interactions in the dataset are sparse, it is difficult to accurately estimate the Q value of every action. Therefore we learn a value function only to estimate the long-term engagement. To approximate the optimal value function in the dataset, we use expectile
regression to yield a asymmetric  $\ell_2$ loss like \cite{PGIQL,VEM,PrefRec}.

\begin{equation}
\label{Vloss}
\mathbb{E}_{\left(s,a, r, s^{\prime}\right) \sim \mathcal{D}}\left[L_2^\tau\left(r(s,a)+\gamma V_{\phi^{\prime}}\left(s^{\prime}\right)-V_\phi(s)\right)\right], 
\end{equation}


where $\ell_2$  $L_2^\tau(u)=|\tau- \mathbb{I}(u<0)| u^2$), $V_\phi(s)$ is a parameterized value function, $V_{\phi^{\prime}}$ is the target value function and can be updated via Polyak averaging \cite{BCQ}. 
Different from DQN TD error\cite{DQN}, the training process of $V_\phi(s)$ only uses $(s,a,r,s')$ in the dataset and there is no $\operatorname{argmax}$ operation for the TD goal $r+\gamma V_{\phi^{\prime}}\left(s^{\prime}\right)$, hence the estimation is conservative. 
In this way, it alleviates the  problem of the overestimation issue mentioned and we will discuss more about the value function loss in the section \ref{experiment}.
Another advantage of only using value function compared with DQN methods is that it does not need to allocate weights and storage to record the Q value of every action, while the action space is very large in the recommendation. 


 
\subsection{Policy Extraction}
After learning the state value function, we turn to the exact policy. 
There exist two key problems here: 1) How to solve the problem of data sparsity and lack of negative feedback; and 2) How to combine the learned state value function to obtain a better policy. 
For the first problem, we design two auxiliary tasks. In particular, we learn two predictive models to predict the reward and next state using the current state-action pair. In addition, we construct positive and negative state-action pairs to learn the model with a new contrastive learning method, where the negative actions are randomly sampled from the uninteracted item space.
For the second problem, we extract the policy by a value-weighted regression method like \cite{IQL}.
We discuss the details in the following.

\subsubsection{\textbf{Reward and State Transition Learning}}

\par
Reward functions and state transition (\textit{i.e.} state transitions) are important elements of MDP. Prior studies \cite{VRL3,VRL5} on visual RL have demonstrated that state representation can be improved by learning state transitions and reward functions as auxiliary tasks.
Since both SeqRec and visual RL have high dimensional discrete state space, learning state transition, and reward functions in SeqRec can also benefit the RL recommender in maintaining long-term profits.
A popular method is learning a bisimulation metric representation in latent space \cite{VRL2},
\begin{equation}
d\left(\mathbf{s}_i, \mathbf{s}_j\right)=\max _{\mathbf{a} \in \mathcal{A}} (1-c)\left|{r}({\mathbf{s}_i},{\mathbf{a}})-{r}({\mathbf{s}_j},{\mathbf{a}})\right|+ c \cdot W_1\left({p}({\mathbf{s}_i},{\mathbf{a}}), {p}({\mathbf{s}_j},{\mathbf{a}}) ; d\right),
\end{equation}
where $W_1$ is Wasserstein metric. Many studies are based on state similarity measurement like bisimulation, and through data augmentation of the state, similar and dissimilar state pairs are constructed for contrastive learning. 

However, given the importance of negative feedback in recommendation systems, we propose a new contrastive learning style that has not yet been studied.  Specifically, we can construct positive and negative pairs of state-action samples for the reward function and state transition based on the negative action samples,
which has been proven to be of vital importance in the recommendation system. Here, for a state $s$, the positive action $a$ is just the action in the dataset, \textit{i.e.}, the item that the user interacts with.
A negative action $a^-$ is defined as a random sample from items that are not interacted with by users.
Intuitively, positive state-action pairs lead to a reward signal and the next state in the dataset, whereas negative state-action pairs do not.
	

\begin{itemize}[leftmargin=10pt]
    \item \textbf{Reward function learning.} 
Given a state and an item interacted with by the user, we can use the corresponding reward signal to learning a reward function. We can maximize the similarity of predictive reward and given reward for positive state-action pairs and minimize the corresponding similarity for the negative state-action pairs. For reward function learning, classical InfoNCE\cite{info} may not be suitable because the scalar reward signal makes it difficult to determine similarity. This process can be formulated as a classification task, where $r$ of negative action is set as $[1,0,0]$ and that of click and purchase is set as $[0,1,0]$ and $[0,0,1]$ respectively. We denote the contrastive loss as  $L_r$,
\begin{equation}
\label{rloss}
L_r = \mathbb{E}_{\left(s, a,s^{\prime}\right)} \left [ {d}\left(\hat{r}\left(z, a\right), r\left(z, a\right)\right)  + \sum_i {d}(\hat{r}(z, a_i^-), r(z, a_i^-))\right ]
\end{equation}

where $z$ is a representation of state $s$ by a sequential model $G(\cdot)$, $\hat{r}$ is a parameterized reward function, and ${d}$ is a cross-entropy loss function. In addition, the loss can be pair-wise reweighed by the reward to distinguish the importance of different state-action pairs.

\item \textbf{State transition learning.}
For state transition learning, we want to capture the relationship that the latent vector in the current state will transfer to the latent vector in the next state when the agent takes an action. 
The states and actions in the dataset give a positive state-action pair for predicting the next state. 
For negative state-action pairs, we conduct negative sampling from the item set that is not interacted with by the user.
Specifically, we use InfoNCE to define state transition learning loss,
\begin{equation}
\label{ploss}
L_p = \mathbb{E}_{\left(s, a,r\right)} \left [ - \log \left(\frac{e^{\operatorname{sim}(\hat{p}(z, a), z') / \tau}}{ e^{\operatorname{sim}\left(\hat{p}\left(z, a\right), z'\right) / \tau}+\sum_i e^{\operatorname{sim}(\hat{p}(z, a_i^-), z') / \tau}}\right) \right ],
\end{equation}
where $z,z'$ is representation of $s,s'$, $\hat{p}$ is a  parameterized state transition function, $\tau$ is a temperature factor and $\operatorname{sim}$ is defined as cosine similarity. Similarly in reward function learning, we pair-wise reweigh the similarity of negative state-action pairs and the exponential value of similarity of positive state-action pairs in the numerator with reward. In this way, we need not fine-tune the temperature factor $\tau$ , which is usually sensitive in contrastive learning. 
\end{itemize}

 
\subsubsection{\textbf{Algorithm Summary}}
After learning a value function, we next extract the recommendation policy. Like \cite{IQL,PGIQL}, we use the value-weighted regression method to extract the policy, which is an effective way to implement RL in offline setting. Considering the simplicity and efficiency, the policy extract loss $L_{\pi}$ is defined as
\begin{equation}
\label{piloss}
L_\pi=\mathbb{E}_{(s, a, r,s') \sim \mathcal{D}}\left[ - (r(s,a) + \gamma  V_{{\phi^'}}(s') )\log \pi(a \mid s)\right],
\end{equation}
where $V_{{\phi^'}}$ is the target value function got by \eqref{Vloss} and $\pi_{\theta}(a\mid s)$  is the recommendation policy.  Since we have learned the reward function and state transition, \eqref{ploss} and \eqref{rloss} can be added to \eqref{piloss} as auxiliary tasks to enhance the representation, and we can get the loss of the whole policy extraction network,
\begin{equation}
\label{Ploss}
L_P(\theta)= \alpha L_{\pi} +  L_r +  L_p,
\end{equation}
where $\alpha$ is a coefficient that balances different losses.

In general, the training process of the whole algorithm MCRL can be summarized as Algorithm \ref{alg1}.

\section{EXPERIMENTS}
\label{experiment}
In this section, we evaluate the proposed MCRL in the e-commerce scenario on two real-world datasets to demonstrate its superiority  and reveal the reasons for its effectiveness by answering the following research questions:

\begin{itemize}[leftmargin=*]
    \item \textbf{RQ1:} How does MCRL perform compared with state-of-the-art supervised methods, offline RL methods, and self-supervised RL methods?   
    \item \textbf{RQ2:} How do different components (\textit{i.e.}, value function, reward model, and state transition model) affect the effectiveness of the proposed MCRL?
    \item \textbf{RQ3:} How do contrastive learning settings affect the effectiveness of the proposed MCRL?
    \item \textbf{RQ4:} How does the number of negative actions and discount factor influence the performance?
\end{itemize}

\subsection{Dataset Description}
To evaluate the effectiveness of MCRL, we conduct experiments on two public datasets: $\mathbf{RetailRocket}$\footnote{https://www.kaggle.com/datasets/retailrocket/ecommerce-dataset} and $\mathbf{Yoochoose}$\footnote{https://recsys.acm.org/recsys15/challenge/}.
Table \ref{tab:dataStatics} presents these datasets’ detailed statistics:
\begin{itemize}[leftmargin=*]
    \item $\mathbf{RetailRocket}$ dataset is collected from a real-world e-commerce website, where sequential data of user’s behaviour of  viewing and adding items to a shopping cart is recorded. For simplicity,  views and adding to a cart are regarder as clicks and purchases. We remove the items which are interacted by the user less than 3 times. Sequences whose length is shorter than 3  are also removed. There are 1,176,680 clicks and 57,269 purchases over 70,852 items in the processed dataset.

    \item $\mathbf{Yoochoose}$ dataset is from RecSys Challenge 2015. 
    Each user-item interaction session consists of a sequence of user clicks or purchases. We remove sessions that are shorter than 3 items. We randomly sample from the 200k sessions to obtain a dataset containing 1,110,965 clicks and 43,946 purchases upon 26,702 items.

\end{itemize}

Table 1 presents these datasets’ detailed statistics.

\subsection{Experimental Settings}
\subsubsection{\textbf{Evaluation protocols}}

Following previous work \cite{xin}, we divide each dataset into training, validation, and test sets at a ratio of 8:1:1 and use cross-validation to evaluate the proposed methods. For validation and test sets, the events of a sequence are provided one by one, and then we check the rank of the item of the next event. We rank the recommended item among the whole item set. We repeat each experiment 3 times to report the average performance.

 Two widely used metrics hit ratio (HR) and normalized discounted cumulative gain (NDCG) are adopted to evaluate the effectiveness of MCRL. In SeqRec, HR@K is equivalent to Recall@K, measuring whether the ground-truth item is in the top-k positions of the recommendation list. In this work, HR@K for click is defined as 
\begin{equation}
\label{hr}
\text{HR(click)} = \frac{\#\text{hit among clicks}}{\#\text{clicks in the test}}.
\end{equation}
HR(purchase)@K is defined similarly. NDCG is sensitive to rank position, which assigns higher scores to top positions in the recommendation list \cite{NDCG}. We set K as 5,10,20 in the reported results.

In the e-commerce scenario, the recommendation system is more interested in increasing the number of purchases than clicks, hence we assign a higher reward to actions leading to purchases.
For the recommended item that is not interacted with by the user, a zero reward is assigned and we treat it as a negative action. Therefore, the cumulative reward (\textit{i.e.}, long-term engagement) is proportional to HR and NG \cite{xin2}. 

 \subsubsection{\textbf{backbone networks}} 
We adopt three typical SeqRec methods as the backbone networks when offline RL and self-supervised RL methods are implemented, similar to \cite{xin,xin2,prl,CSA}.

\begin{itemize}[leftmargin=*]
    \item $\mathbf{GRU}$ \cite{GRU4Rec}: This is an RNN-based sequential model and leverages GRU to encode the input user-item interaction sequence.
    \item $\mathbf{Caser}$ \cite{Caser}: This is a recently proposed CNN-based sequential recommender that applies convolution operations on the embedding matrix of the input sequence.
    \item $\mathbf{SASRec}$ \cite{SASRec}: This is an attention-based sequential recommender and uses a Transformer encoder to encode the input sequence.
\end{itemize}

 \subsubsection{\textbf{Baselines}} 
 To demonstrate the effectiveness, we compare our proposed MCRL with different offline RL, and self-supervised RL methods with different backbone networks. Due to the unavailability of open source code, some related work will be discussed further in section \ref{reltaedwork}.

\begin{itemize}[leftmargin=*]
    \item $\mathbf{IQL}$ \cite{IQL}: This is a state-of-the-art offline RL method and takes a state conditional upper expected to estimate the value of the best actions in a state.
    \item $\mathbf{SQN}$ \cite{xin}: This is a representative self-supervised RL method that uses an RL head to enhance the representation.
    
    \item $\mathbf{SAC}$ \cite{xin}: This is also a self-supervised RL method and utilizes these Q-values to re-weight the supervised part.

    \item  $\mathbf{CDARL}$  \cite{CDARL}: This is a self-supervised RL method considering data augmentation and intrinsic reward. 
    
\end{itemize}

\begin{table*}[t]
\caption{Top-k performance comparison of supervised, offline RL, and self-supervised RL methods and MCRL on RetailRocket dataset$ (k = 5, 10, 20)$. 'HR’ and ’NG’  denotes hit ratio and NDCG respectively. Recommendations are generated from the policy $\pi$. Boldface denotes the highest score.}
\vspace{0.3px}
\label{tab:res_erm}
\renewcommand\arraystretch{1.0}
\begin{tabular}{cccccccccccccc}
\hline
{\multirow{2}{*}{Models}}&&&\multicolumn{2}{c}{Purchase}&&&&&\multicolumn{2}{c}{Click}&&\\  \cmidrule(r){2-7} \cmidrule(r){8-14}
 &HR@5&NG@5&HR@10&NG@10&HR@20&NG@20 &HR@5&NG@5&HR@10&NG@10&HR@20&NG@20& \\
\hline
GRU 
&0.4608&0.3834&0.5107&0.3995&0.5564&0.4111  &0.2233&0.1735&0.2673&0.1878&0.3082&0.1981\\\cdashline{2-14}
GRU-IQL  
&0.5321&0.4451&0.5858&0.4624&0.6328&0.4743  &0.2654&0.2056&0.3181&0.2226&0.3659&0.2348\\\cdashline{2-14}
GRU-SQN   
&0.5069&0.4130&0.5589&0.4289&0.5946&0.4392  &0.2487&0.1939&0.2967&0.2094&0.3406&0.2205\\
GRU-SAC  
&0.4942&0.4179&0.5464&0.4341&0.5870&0.4428  &0.2451&0.1924&0.2930&0.2074&0.3371&0.2186\\\cdashline{2-14}

GRU-CDARL  
&0.5882&0.5047&0.6371&0.5205&0.6730&0.5296  &0.2921&0.2277&0.3482&0.2459&0.4004&0.2591\\\cdashline{2-14}

GRU-MCRL 
&\textbf{0.5972}&\textbf{0.5071}&\textbf{0.6481}&\textbf{0.5229}&\textbf{0.6900}&\textbf{0.5339}  &\textbf{0.2779}&\textbf{0.2151}&\textbf{0.3301}&\textbf{0.2326}&\textbf{0.3806}&\textbf{0.2455}\\
\hline

Caser 
&0.3491&0.2935&0.3857&0.3053&0.4198&0.3141  &0.1966&0.1566&0.2302&0.1675&0.2628&0.1758\\\cdashline{2-14}
Caser-IQL  
&0.3855&0.3288&0.4266&0.3420&0.4618&0.3509  &0.2125&0.1704&0.2504&0.1827&0.2854&0.1916\\\cdashline{2-14}
Caser-SQN   
&0.3674&0.3089&0.4050&0.3210&0.4409&0.3301  &0.2089&0.1661&0.2454&0.1778&0.2803&0.1867\\
Caser-SAC  
&0.3871&0.3234&0.4436&0.3386&0.4763&0.3494  &0.2206&0.1732&0.2617&0.1865&0.2999&0.1961\\\cdashline{2-14}
Caser-CDARL  
&0.4073&0.3438&0.4521&0.3563&0.4822&0.3655  &0.2153&0.1716&0.2532&0.1838&0.2879&0.1924\\\cdashline{2-14}
Caser-MCRL 
&\textbf{0.4877}&\textbf{0.4171}&\textbf{0.5387}&\textbf{0.4337}&\textbf{0.5851}&\textbf{0.4455}  &\textbf{0.2498}&\textbf{0.1979}&\textbf{0.2941}&\textbf{0.2123}&\textbf{0.3374}&\textbf{0.2232}\\
\hline

SASRec 
&0.5267&0.4298&0.5916&0.4510&0.6341&0.4618  &0.2541&0.1931&0.3085&0.2107&0.3570&0.2230\\\cdashline{2-14}
SASRec-IQL  
&0.5403&0.4442&0.6016&0.4643&0.6519&0.4643  &0.2737&0.2084&0.3286&0.2263&0.3813&0.2396\\\cdashline{2-14}
SASRec-SQN   
&0.5681&0.4617&0.6203&0.4806&0.6619&0.4914  &0.2761&0.2104&0.3302&0.2279&0.3803&0.2406\\
SASRec-SAC  
&0.5623&0.4679&0.6127&0.4844&0.6505&0.4940  &0.2670&0.2056&0.3208&0.2230&0.3701&0.2355\\\cdashline{2-14}
SASRec-CDARL
&0.5780&0.4799&0.6315&0.4981&0.6785&0.5088  &0.2806&0.2145&0.3386&0.2327&0.3913&0.2458\\\cdashline{2-14}
SASRec-MCRL 
&\textbf{0.6547}&\textbf{0.5755}&\textbf{0.6967}&\textbf{0.5891}&\textbf{0.7296}&\textbf{0.5974}  &\textbf{0.2815}&\textbf{0.2205}&\textbf{0.3342}&\textbf{0.2375}&\textbf{0.3827}&\textbf{0.2498}\\
\hline
\end{tabular}
\vspace{0.3cm}
\end{table*}

\begin{table*}[t]
\caption{Top-k performance comparison of supervised, offline RL, and self-supervised RL methods and MCRL on Yoochoose dataset$ (k = 5, 10, 20)$. 'HR’ and ’NG’  denotes hit ratio and NDCG respectively. Recommendations are generated from the policy $\pi$. Boldface denotes the highest score.}
\vspace{0.1px}
\label{tab:res_erm}
\renewcommand\arraystretch{1.0}
\begin{tabular}{cccccccccccccc}
\hline
{\multirow{2}{*}{Models}}&&&\multicolumn{2}{c}{Purchase}&&&&&\multicolumn{2}{c}{Click}&&\\  \cmidrule(r){2-7} \cmidrule(r){8-14}
 &HR@5&NG@5&HR@10&NG@10&HR@20&NG@20 &HR@5&NG@5&HR@10&NG@10&HR@20&NG@20& \\
\hline
GRU 
&0.3994&0.2824&0.5183&0.3204&0.6067&0.3429  &0.2876&0.1982&0.3793&0.2279&0.4581&0.2478\\\cdashline{2-14}
GRU-IQL  
&0.4220&0.3010&0.5350&0.3378&0.6243&0.3606  &0.3213&0.2225&0.4186&0.2541&0.4987&0.2744\\\cdashline{2-14}
GRU-SQN   
&0.4228&0.3016&0.5333&0.3376&0.6233&0.3605  &0.3020&0.2093&0.3946&0.2394&0.4741&0.2587\\
GRU-SAC  
&0.4394&0.3154&0.5525&0.3521&0.6378&0.3739  &0.2863&0.1985&0.3764&0.2277&0.4541&0.2474\\\cdashline{2-14}
GRU-CDARL
&\textbf{0.4726}&\textbf{0.3395}&\textbf{0.5906}&\textbf{0.3779}&\textbf{0.6853}&\textbf{0.4018} 
&{0.3228}&{0.2244}&{0.4198}&{0.2552}&{0.5022}&{0.2752}\\\cdashline{2-14}
GRU-MCRL 
&{0.4588}&{0.3296}&{0.5703}&{0.3660}&{0.6492}&{0.3860}  &\textbf{0.3263}&\textbf{0.2262}&\textbf{0.4220}&\textbf{0.2572}&\textbf{0.5043}&\textbf{0.2780}\\
\hline

Caser 
&0.4475&0.3211&0.5559&0.3565&0.6393&0.3775  &0.2728&0.1896&0.3593&0.2177&0.4371&0.2372\\\cdashline{2-14}
Caser-IQL  
&0.4690&0.3402&0.5756&0.3750&0.6601&0.3964 &0.3017&0.2105&0.3910&0.2394&0.4699&0.2595\\\cdashline{2-14}
Caser-SQN   
&0.4552&0.3302&0.5637&0.3653&0.6417&0.3862  &0.2742&0.1909&0.3613&0.2192&0.4381&0.2386\\
Caser-SAC  
&\textbf{0.4866}&\textbf{0.3527}&\textbf{0.5914}&\textbf{0.3868}&\textbf{0.6689}&\textbf{0.4065}  &0.2726&0.1894&0.3580&0.2171&0.4340&0.2362\\\cdashline{2-14}
Caser-CDARL
&{0.4571}&{0.3295}&{0.5667}&{0.3648}&{0.6442}&{0.3847}  &\textbf{0.3164}&0.2199&\textbf{0.4105}&\textbf{0.2506}&0.4878&0.2702\\\cdashline{2-14}
Caser-MCRL 
&0.4741&0.3441&0.5774&0.3775&0.6622&0.3989  &{0.3159}&\textbf{0.2200}&{0.4095}&{0.2503}&\textbf{0.4894}&\textbf{0.2706}\\
\hline

SASRec 
&0.4228&0.2938&0.5418&0.3326&0.6329&0.3558  &0.3187&0.2200&0.4164&0.2515&0.4974&0.2720\\\cdashline{2-14}
SASRec-IQL  
&0.4459&0.3179&0.5687&0.3579&0.6522&0.3792  &0.3258&0.2241&0.4254&0.2564&0.5079&0.2773\\\cdashline{2-14}
SASRec-SQN   
&0.4336&0.3067&0.5505&0.3435&0.6442&0.3674  &0.3272&0.2263&0.4255&0.2580&0.5066&0.2786\\
SASRec-SAC  
&0.4540&0.3246&0.5701&0.3623&0.6576&0.3846  &0.3130&0.2161&0.4114&0.2480&0.4945&0.2691\\\cdashline{2-14}
SASRec-CDARL
&0.4521&0.3172&0.5667&0.3555&0.6558&0.3783  &0.3282&0.2260&0.4274&0.2584&0.5097&0.2792\\\cdashline{2-14}

SASRec-MCRL 
&\textbf{0.4625}&\textbf{0.3297}&\textbf{0.5756}&\textbf{0.3664}&\textbf{0.6610}&\textbf{0.3881}  &\textbf{0.3345}&\textbf{0.2326}&\textbf{0.4319}&\textbf{0.2642}&\textbf{0.5112}&\textbf{0.2843}\\
\hline
\end{tabular}
\vspace{0.5cm}
\end{table*}

\subsubsection{\textbf{Parameter Settings}}
We implement all methods with Python 3.9.13 and PyTorch 1.9.0 in Nvidia GeForce RTX 3090. For both datasets, the input sequences
are defined as the last 10 items before the target item. We complement the sequence whose length is less than 10 with a padding
item. To ensure a fair comparison, we fix the embedding size of all models to 64. We optimize all models with Adam optimizer \cite{adam} and fix the batch size at 256. The learning rate is set as 0.005 for both Yoochoose and RetailRocket. For GRU, we set the size of the hidden state as 64. For Caser, 1 vertical convolution filter and 16 horizontal filters are used. The height of filters is searched in the range of $\{2,3,4\}$
and the dropout ratio is set as 0.1. For SASRec, we set the number of heads in self-attention as 1. 

For RL-based methods, as recommended by \cite{xin}, we set the discount factor as 0.5 and the ratio between the click reward ($r_c$) and the purchase reward ($r_p$) as  0.2 and 1 respectively.  For negative sampling, a uniform distribution strategy is used. Without special mention, the negative number is set as 30. The temperature factor is set as $1$ and coefficients $\alpha$ can is set as $1$ except in the case of GRU backbone on Yoochoos dataset. When IQL and MCRL are integrated with a base model, the hyperparameter setting remains exactly unchanged for a fair comparison. 

For the implementation of the reward model and state transition model,  after concatenating the embedding of state and action from the supervised backbone network, the representation vector is fed into two fully-connected networks with ReLu as an activation function. The difference between the reward model and the state transition model is the last layer, where the output dims are 3 and 64 respectively. For the value function, its structure is similar to the reward model except that the output dim is 1.

\begin{figure*}[htbp] 
\begin{minipage}[t]{0.34\linewidth} 
\centering
\includegraphics[width=2.3in, height=1.6in]{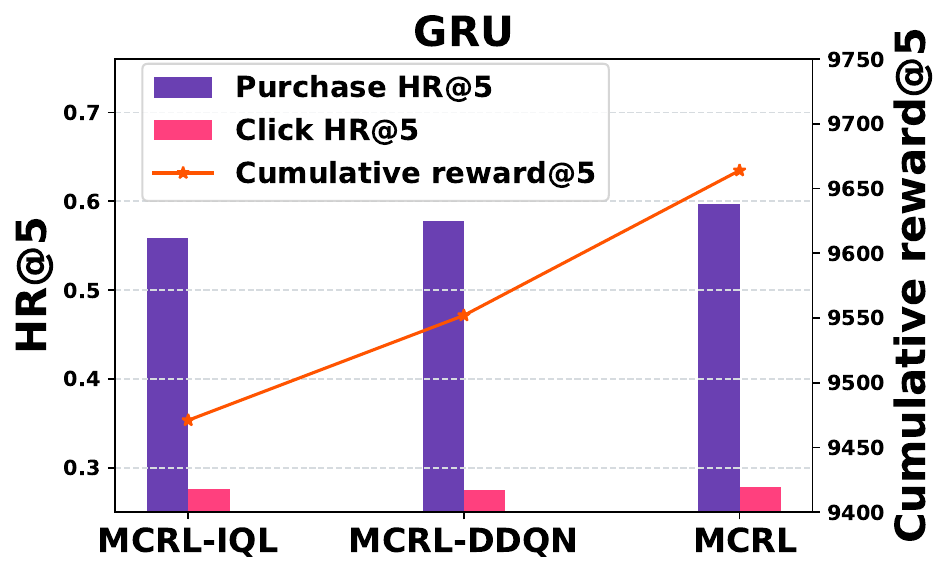} 
\end{minipage}%
\begin{minipage}[t]{0.34\linewidth}
\centering
\includegraphics[width=2.3in, height=1.6in]{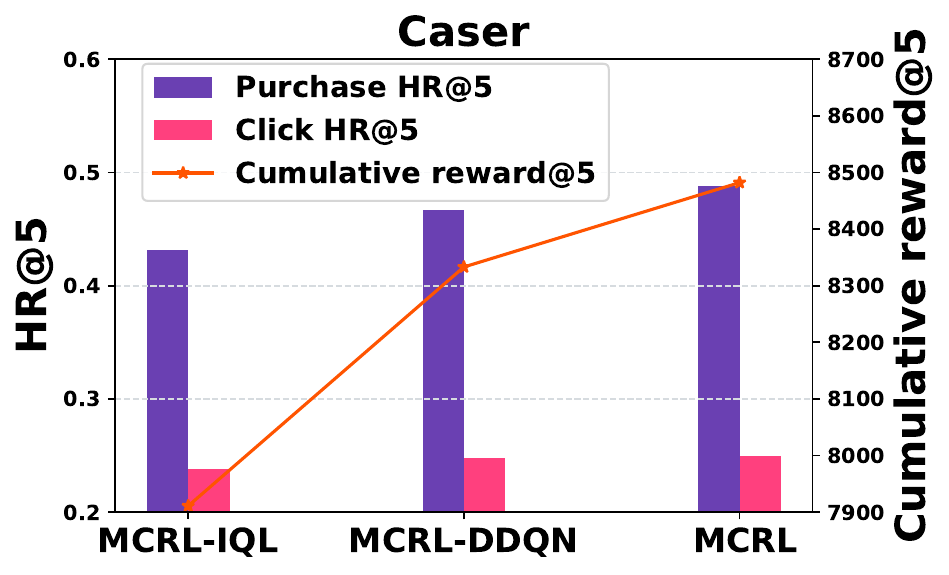}
\end{minipage}%
\begin{minipage}[t]{0.34\linewidth}
\centering
\includegraphics[width=2.3in, height=1.6in]{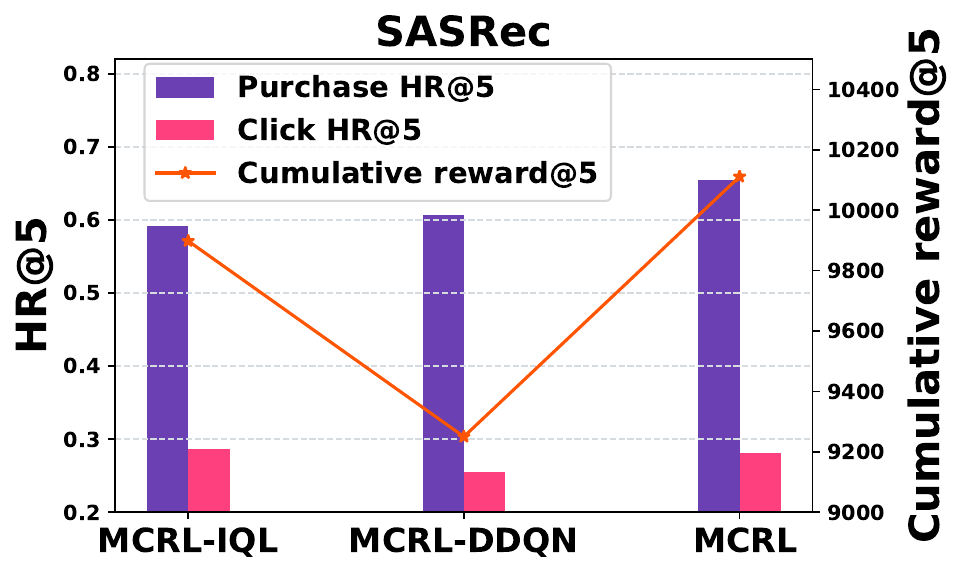}
\vspace{-0.5cm}
\end{minipage}
\caption{The study of the value function module on RetailRocket. }
\label{valueablation}
\end{figure*}

\begin{figure*}[htbp] 
\begin{minipage}[t]{0.33\linewidth} 
\centering
\includegraphics[width=2.3in, height=1.6in]{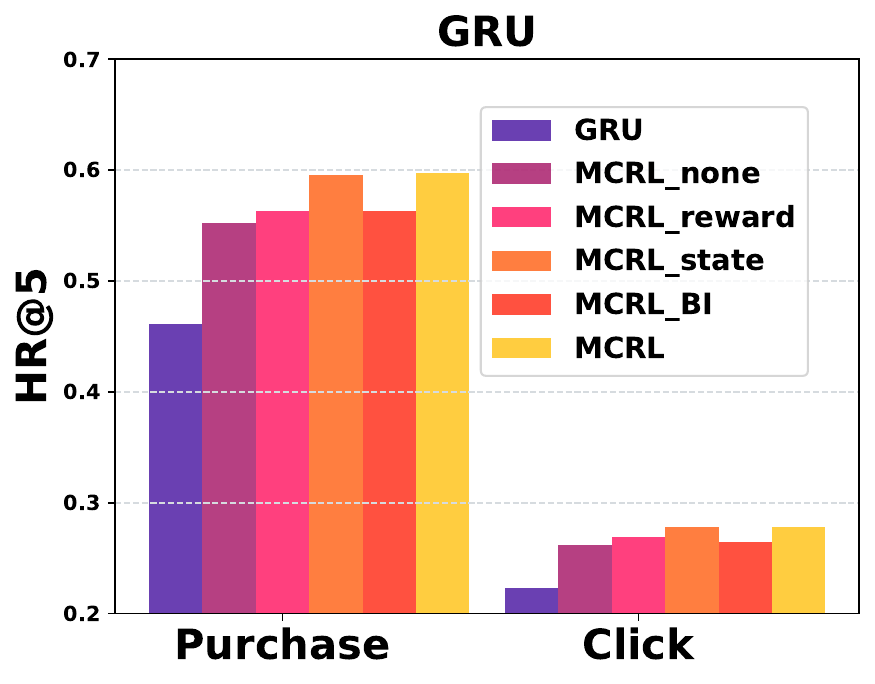} 
\end{minipage}%
\begin{minipage}[t]{0.33\linewidth}
\centering
\includegraphics[width=2.3in, height=1.6in]{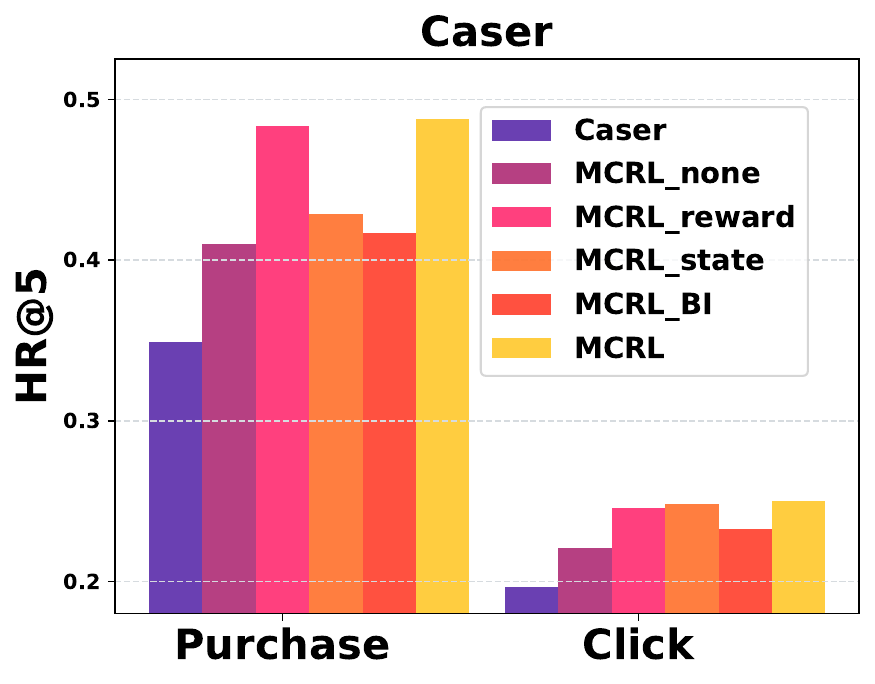}
\end{minipage}%
\begin{minipage}[t]{0.33\linewidth}
\centering
\includegraphics[width=2.3in, height=1.6in]{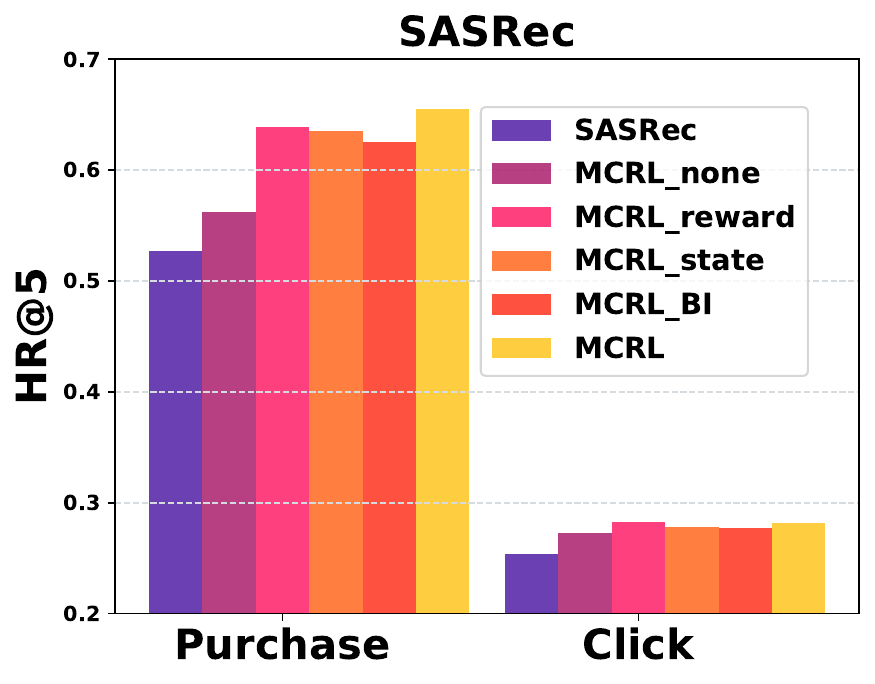}
\vspace{-0.5cm}
\end{minipage}
\caption{The study of reward and state transition model for MCRL on RetailRocket}
\label{Modelablation}
\end{figure*}

\subsection{Performance Comparison (RQ1)}
Table 2 and Table 3 illustrate the performance of top-k recommendation
on RetailRocket and Yoochoose, respectively.

On the RetailRocket dataset, the proposed achieves the highest scores in all situations for purchase and click, which demonstrates the items recommended by MCRL have a higher probability to be clicked or purchased by users. Notice that MCRL achieves the best performance for both HR and NDCG, indicating that MCRL tends to give the items that have a higher purchase reward with a higher ranking position in the recommendation list. Compared with the supervised baseline, IQL, SQN, SAC,CDARL, and MCRL all have better performance, which shows the potential of applying RL to the SeqRec. In addition, the value function estimates the long-term engagement to improve the policy learning, hence the improvement in purchase is more obvious in the e-commerce scenario. In comparison with baseline RL-based methods, MCRL still achieves significant performance improvement for all backbone networks. It demonstrates that MCRL effectively improves the RL-based recommendation model with the help of exploiting MDP structures.  

On the Yoochoose dataset, we observe that MCRL outperforms all supervised methods in terms of click and purchase except for the purchase prediction when integrating with the Caser model, where SAC has the highest scores.
However, SAC and MCRL both outperform the supervised methods, IQL and SQN, and their performance gap is small.
Despite the improvement of MCRL over almost all baselines, the score increase is smaller compared to that on RetailRocket.
Regarding the larger improvement of MCRL on RetailRocket compared to Yoochoose, we
believe that the signal sparsity problem is more severe in RetailRocket due to its larger number of items(70,852 in RetailRocket and 26,702 in Yoochoose) and hence larger state and action space. As MCRL is designed to address the signal sparsity problem, it is likely to have a larger impact on datasets with more items.

\subsection{ Study of MCRL (RQ2)}
To get deep insights into the effectiveness of the learned value function, reward model, and state transition model,  we conduct ablation experiments. For space saving, we only report the results on the RetailRocket and the trend for Yoochoose is similar. Besides, we enhance IQL with the learned model to further see the effect of the learned model, since they can also be regarded as a relatively independent part.  


\subsubsection{\textit{ Effect of the value function}}
In this part, we use different RL methods to estimate long-term engagement to verify the rationality of the proposed value function. Besides 
 HR, we also report the cumulative rewards@k, which means the sum of rewards for top-k recommendation in the test set. For space-saving, we do not report results about NDCG, which shows a similar trend. Figure \ref{valueablation} illustrates the comparison of supervised model Gru, MCRL-DQN, MCRL-IQL, and MCRL, where MCRL-DQN are MCRL-IQL are different from MCRL about the Q function learning.  For MCRL-DQN, we implement a double Q learning \cite{doubleQ} which learns two Q functions and use the conservative one, to estimate the expected cumulative rewards. For MCRL-IQL, we use the Q function learning methods proposed in \cite{IQL}, and use an advantage function to learn the policy. MCRL-DDQN performs better than MCRL-IQL for GRU and Caser, but gets a lower click HR, purchase HR and cumulative rewards.  However, the performance of MCRL is better than compared methods for click prediction, purchase prediction, and the cumulative rewards on the test set when instantiated with GRU, Caser, and SASRec backbones, which proves that the proposed value function is more accurate, that is, it can better ease the overestimation problem than baseline offline RL component. Besides, MCRL is superior to them with respect to the number of parameters without directly learning a Q function, that is, the output dim of MCRL's value function is $1$ while that of other RL-based methods is the item size.

\subsubsection{\textit{ Effect of reward and state transition model for MCRL and IQL}}
We vary the setting of MCRL to investigate the impact of learned reward and state transition model. For simplicity's sake, we only report HR metrics since NDCG shows a similar trend. Figure \ref{Modelablation} gives the corresponding results. MCRL-none is the variant of MCRL without learning the reward model and state transition model. MCRL-reward and MCRL-state are MCRL variants with only reward model and state transition model, respectively. For GRU, Caser, and SASRec, all methods with the model gain performance improvement, which shows the effectiveness of the proposed model learning. However, the combination of two models outperforms any single model in all cases, indicating that it is necessary to learn the two models. Moreover, for GRU, the state transition model is more effective than the reward model. For Caser and SASRec,  the conclusion is the opposite, which indicates the different capacities of the two proposed models for different backbone networks.  To further understand the role of model learning, we also implement bisimulation(see section \textit{3.2.1}) using MCRL framework, i.e. MCRL-BI, which learns the reward and transition simultaneously. The result shows that it can also boost the performance since it outperforms MCRL-none for different backbone networks. 
    
\begin{figure*}[htbp] 
\begin{minipage}[t]{0.34\linewidth} 
\centering
\includegraphics[width=2.3in, height=1.6in]{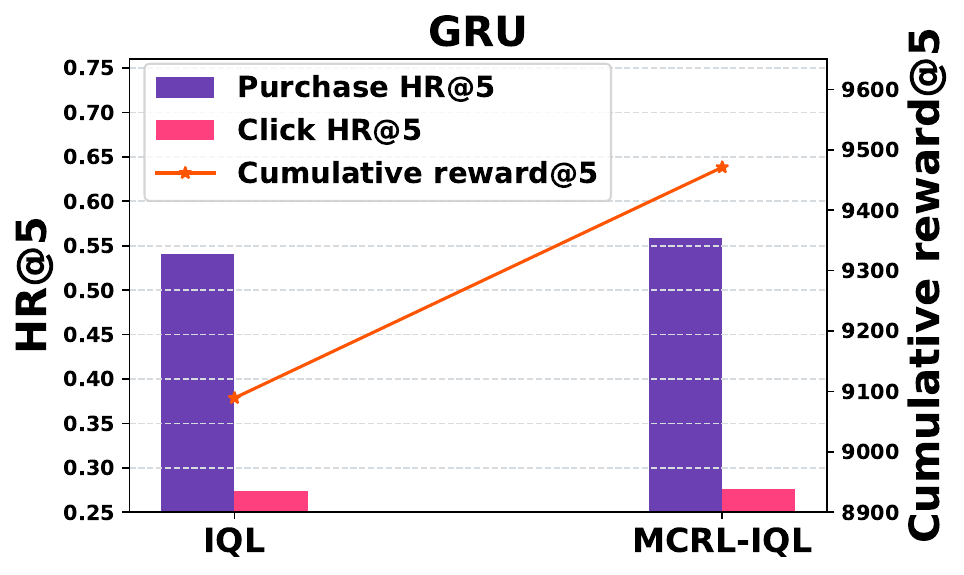} 
\end{minipage}%
\begin{minipage}[t]{0.34\linewidth}
\centering
\includegraphics[width=2.3in, height=1.6in]{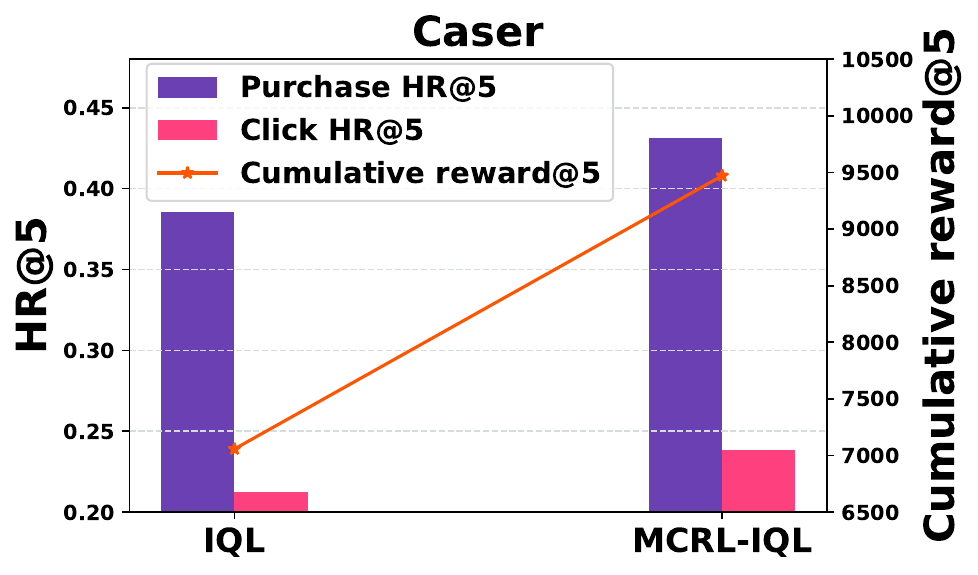}
\end{minipage}%
\begin{minipage}[t]{0.34\linewidth}
\centering
\includegraphics[width=2.3in, height=1.6in]{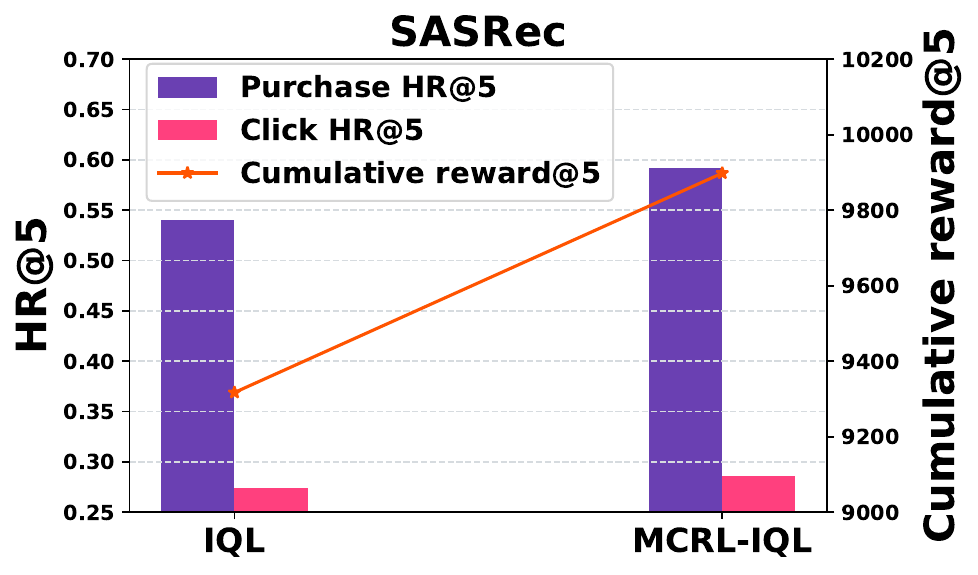}
\vspace{-0.5cm}
\end{minipage}
\caption{Reward and state transition model boost the offline RL method IQL on RetailRocket.}
\label{IQL}
\end{figure*}

We can see that the learned reward and state transition model play an important role in the performance improvement of MCRL and we wonder whether the learned models can also boost other offline RL methods. Figure \ref{IQL} shows that learning a reward model and state transition model (MCRL-IQL) can improve the performance of the base offline RL method (IQL)  for their backbones on RetailRocket with respect to click HR, purchase HR, and cumulative rewards. 

\subsection{Contrastive Learning Study (RQ3)}
\label{crl}

\begin{figure}[h]
    \centering
    \hspace{-10mm}
    \begin{subfigure}[b]{0.2\textwidth}
        \centering
        \includegraphics[width=4.5cm]{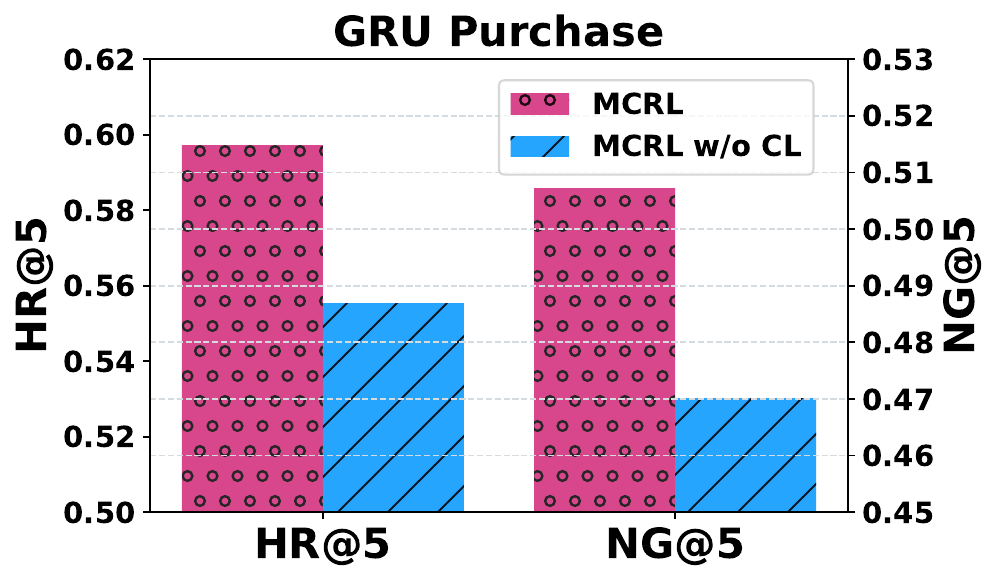}
    \end{subfigure}
    \hspace{8mm}
    \begin{subfigure}[b]{0.2\textwidth}
        \centering
        \includegraphics[width=4.5cm]{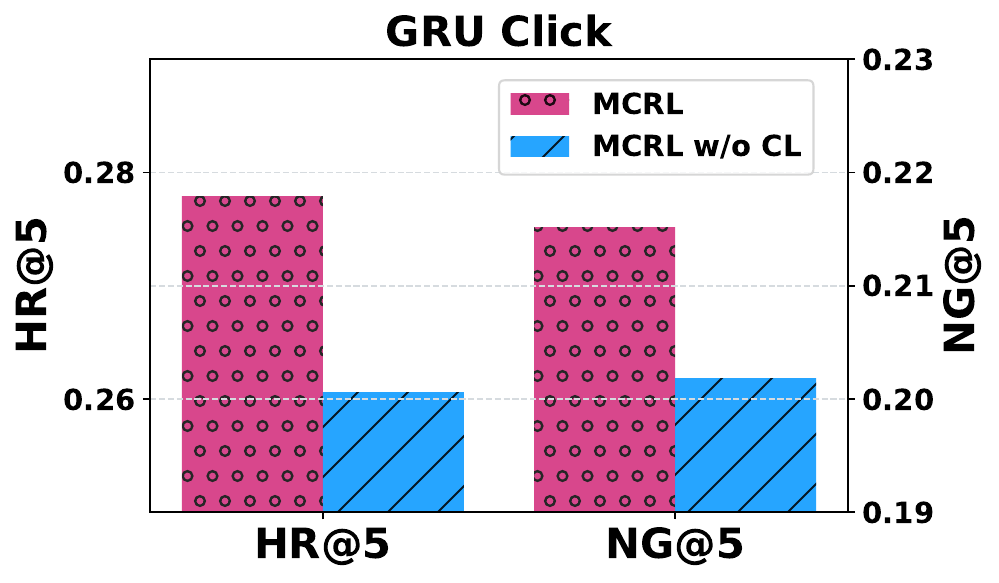}
    \end{subfigure}
        
    \hspace{-10mm}
    \begin{subfigure}[b]{0.2\textwidth}
        \centering
        \includegraphics[width=4.5cm]{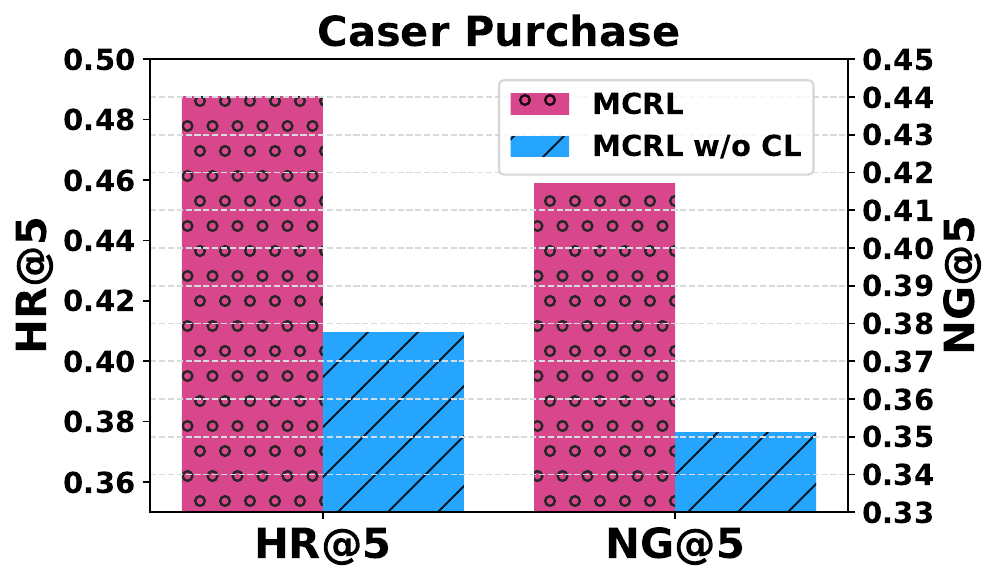}
    \end{subfigure}
    \hspace{8mm}
    \begin{subfigure}[b]{0.2\textwidth}
        \centering
        \includegraphics[width=4.5cm]{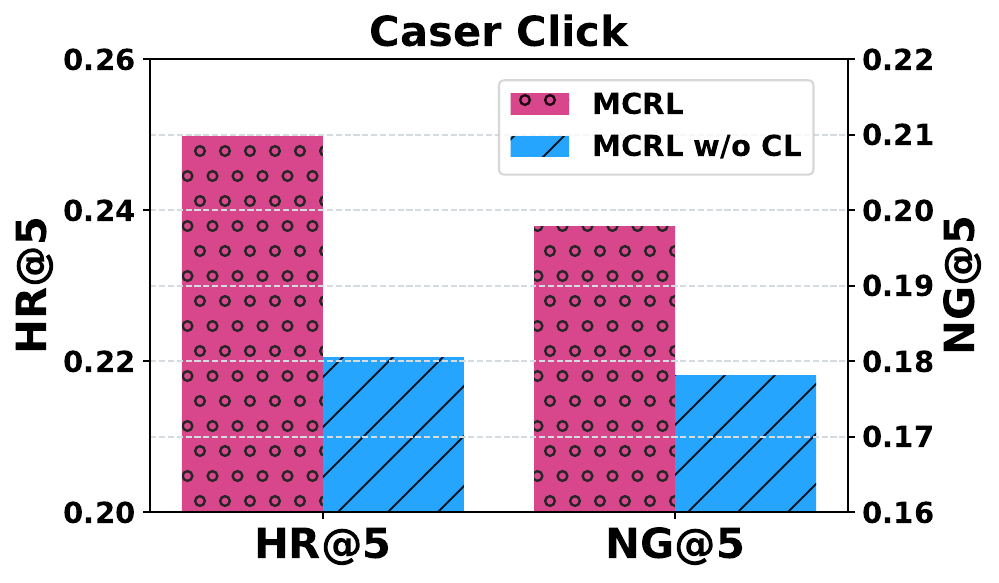}
    \end{subfigure}

    \hspace{-10mm}
    \begin{subfigure}[b]{0.2\textwidth}
        \centering
        \includegraphics[width=4.5cm]{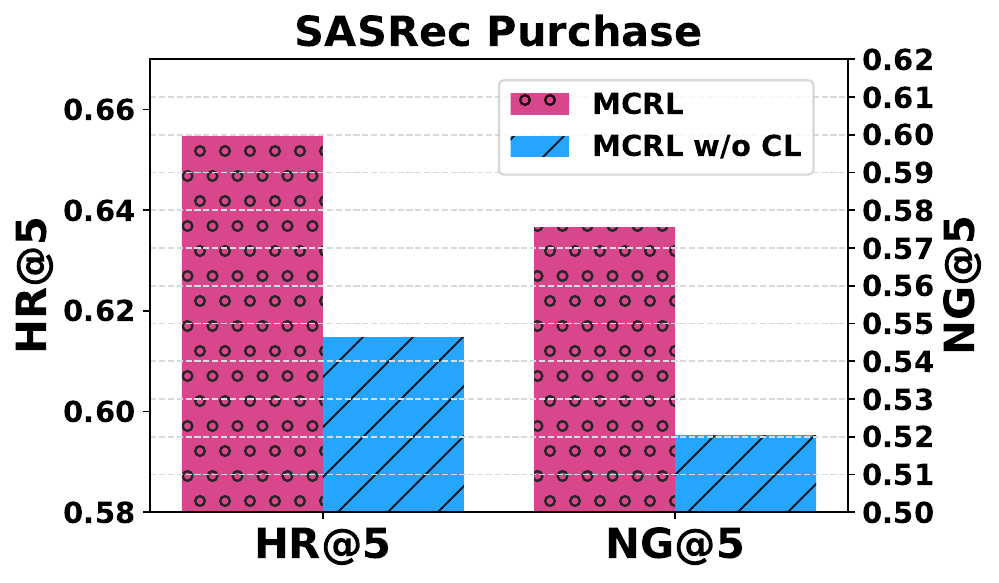}
    \end{subfigure}
    \hspace{8mm}
    \begin{subfigure}[b]{0.2\textwidth}
        \centering
        \includegraphics[width=4.5cm]{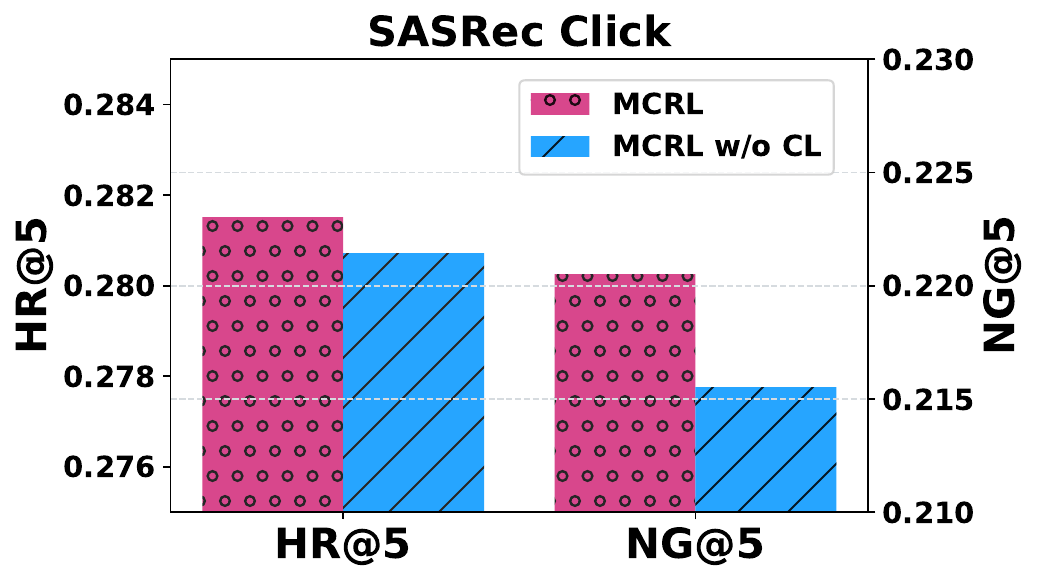}
    \end{subfigure}
    \caption{The ablation study of contrastive learning on RetailRocket.}
    \label{contrastive}
\end{figure}

In this section, we study the effect of contrastive learning. Figure \ref{contrastive} shows the comparison of MCRL and MCRL w/o CL, which learns a reward and state transition model only with positive state-action pairs. On the RetailRocket dataset, MCRL outperforms MCRL w/o CL with respect to HR and NDCG in the prediction of both click and purchase, showing the effectiveness of contrastive learning.  It gives us the insight that utilizing negative reward can ease the data sparsity problem for offline RL learning in recommendation. 


\subsection{Hyperparameter Study (RQ4)}

In this part, we investigate how the hyperparameter setting of the number of negative actions and discount factor influence the performance MCRL. We only report the result of HR@5 for GRU backbone on RetailRocket, since the patterns in other cases are similar.

\subsubsection{\textit{Effect of Discount Factor}}

Figure \ref{Hyperparam} illustrates the influence of the discount factor. From the results of RetailRocket, we see that the performance fluctuates with the discount factor varying and the tendency for click prediction and purchase prediction is similar. However, the fluctuation range are both small (less than \%3 ) for click HR and purchase HR. Overall, the proposed method is not sensitive to the discount factor. 

\subsubsection{\textit{Effect of the Number of Negative actions}}

\begin{figure}
    \centering
    \hspace{-10mm}
    \begin{subfigure}[b]{0.2\textwidth}
        \centering
        \includegraphics[width=4.5cm]{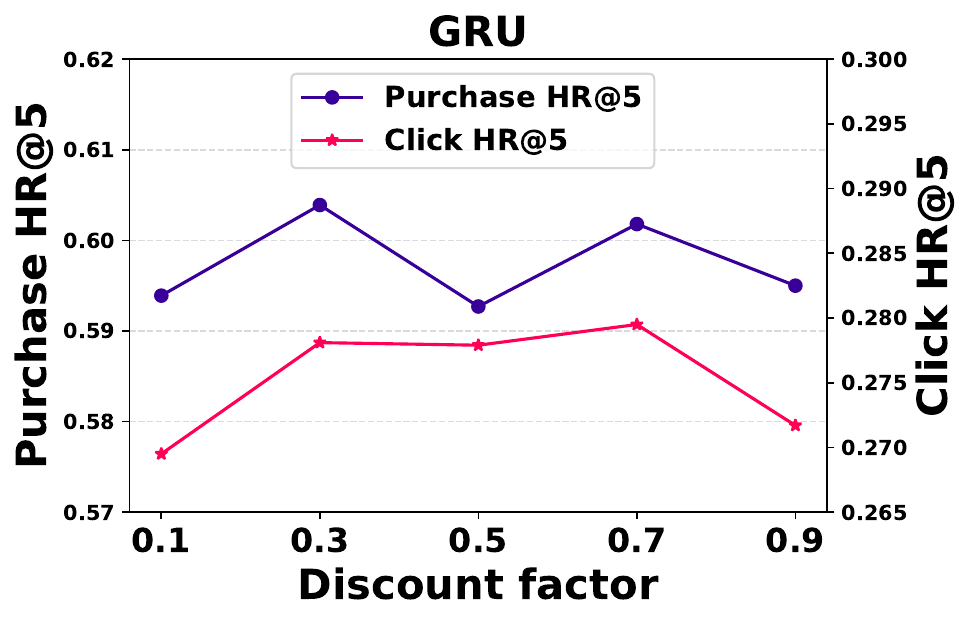}
    \end{subfigure}
    \hspace{8mm}
    \begin{subfigure}[b]{0.2\textwidth}
        \centering
        \includegraphics[width=4.5cm]{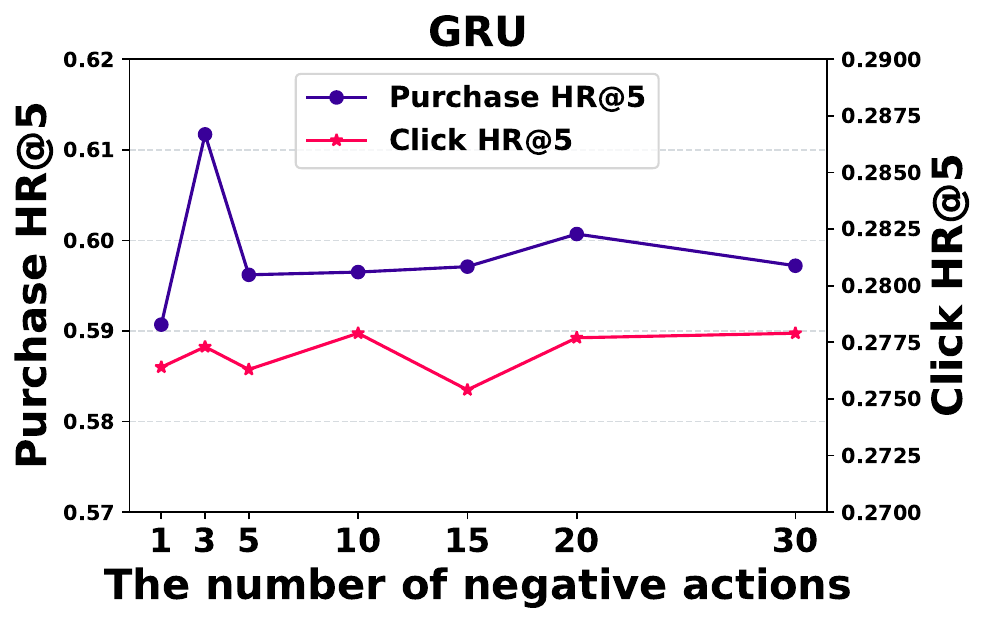}
    \end{subfigure}
    \caption{Hyperparameter Study of discount factor and the number of negative actions on RetailRocket.}
    \label{Hyperparam}
\end{figure}

Figure \ref{Hyperparam} shows the influence of the number of negative actions. The performance remains nearly unchanged when there are more than $5$ negative actions. In conclusion, our contrastive learning module shows robustness with respect to the number of negative actions.

\subsubsection{\textit{Effect of the Training Steps}}
\begin{figure}
    \centering
    \includegraphics[width=0.7\linewidth]{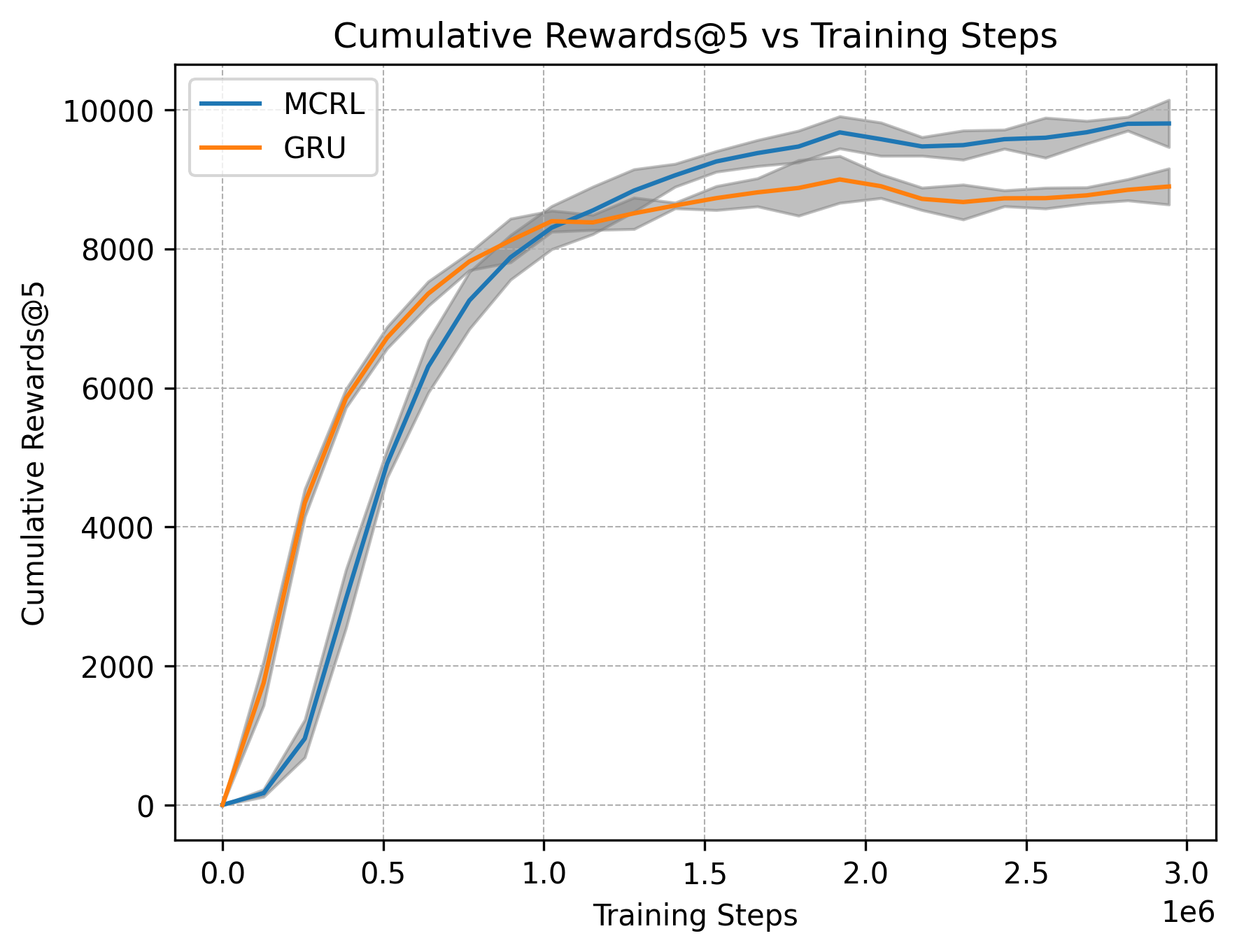}
    \caption{Learning curves of supervised and RL recommender, averaged over 5 runs. }
    \label{learningcurve}
\end{figure}

From Figure \ref{learningcurve}, we can see that MCRL is not as effective as the supervised methods when the number of training steps is small, possibly due to the higher training loss and slower convergence. As the number of training steps increases, MCRL continuously surpasses the supervised methods, demonstrating the effectiveness and stability of MCRL.

\section{Related work}
\label{reltaedwork}
In this section, we introduce some related tasks: sequential recommendation and RL for recommendation.

\subsection{Sequential Recommendation }
Sequential recommendation utilizes self-supervised learning to predict the next item based on the historical sequence of user interaction. In the early stage, Markov Chains \cite{MC1, MC2} and factorization methods \cite{MC3} were widely used in the sequential recommendation, but are limited in the expressiveness of complex sequential signals. Recently, deep learning has played an increasingly important role in the sequential recommendation. For instance, \cite{GRU4Rec} uses gated recurrent units (GRU) as the backbone to extract the sequential representation. \cite {Caser, NextItem} model the user-item sequence based on the utilization of a convolutional neural network (CNN). With transformer \cite{SASRec, Bert4Rec, LSAN, EETSeqRec, Bert, MAE} gaining popularity in various areas, it is also used in sequential recommendations by \cite{SASRec} in the sequence embedding space. Recently, \cite{CL,GCL} focuses on using data augmentation in sequential recommendation with contrastive learning.  \cite{PARS} proposes an architecture that relies on common patterns as well as individual behaviors to tailor its recommendations for each person. 
\cite{ADA} proposes a new training and inference paradigm, termed as Ada-Ranker, to address the challenges of dynamic online serving.  In this article, we have selected three most typical neural network structures  as backbones, similar to \cite{xin,xin2,prl,CSA}.

\subsection{RL for recommendation} 
\label{offlineRL}
Reinforcement learning has been widely used for recommender systems in recent years\cite{SurveyRL4recommondation,surveyRL4rs,RL4RS}.  
\cite{MDPRS} first proposes to model recommendation as a Markov Decision Process, and a model-based RL method for is designed for book recommendation. \cite{Taobao,RL4RS,kuaisim} study how to use collecting recommendation data to construct the recommender simulator.  
\cite{big} apply RL to recommendation with large discrete action spaces and demonstrates the effectiveness on various recommendation tasks with more than one million actions. 
\cite{Resact} proves that the application of offline RL techniques to sequential recommendation is promising by learning a policy near the behaviour policy. 
\cite{RNF} introduces the impact of negative feedback through learning the Q function. \cite{PrefRec} uses reinforcement learning from human feedback to capture the interest of users. \cite{multitaskrl1,multitaskrl2} adopt RL techniques to the multi-task recommendation.
Self-supervised reinforcement (SSR), as a technique in which the model is trained by itself without external label information, is receiving growing interest in robotics \cite{SSL}. Moreover, \cite{xin} proposes two SSR learning frameworks named  Self-Supervised Q-learning (SQN) and Self-Supervised Actor-Critic (SAC) , utilizing the RL component as a form of a regularizer for the supervised learning component, SQN, and SAC can effectively lead to more purchases than clicks. 
\cite{CDARL,CSA} consider to combine the data augmentation with SSR to boost the state representation.   \cite{prl} considers take the cumulative reward signals of sequence trajectory as the input and reformulates SSR as the task of action prediction. However, all these methods ignore mining the structure information of MDP. 
\section{CONCLUSION AND FUTURE WORK}

In this work, we explore the application of RL to sequential recommendations for optimizing long-term engagement. We devise a new framework MCRL, which implicitly learns a model to enhance policy learning. The core insight of this work is to mine the structural information of the MDP and consider the negative feedback. Extensive experiments on two real-world datasets show the rationality and effectiveness of MCRL.

This work shows the potential of enhancing RL-based sequential recommendation methods with MDP structure information and the importance of utilizing negative feedback, which is combined via contrastive learning. Besides reward function and environment dynamics, other structural information of MDP indeed exists in real-world scenarios, such as reward sequence. For instance, we can learn a model to predict the future reward or state sequence. \cite{offpolicy} introduced an off-policy correction term (propensity
score) for the policy-gradient method to evaluate the off-policy effect, this technique can be considered in MCRL. An simple implementation is to reweigh Eqn. \eqref{piloss} with propensity score and evaluate it with metric in \cite{NDCGoff}. 

\section{Appendix}

\subsection{The Algorithm of MCRL}
Here is the pseudocode for the overall MCRL algorithm.

\begin{algorithm}[t]
    \SetAlgoLined 
	\caption{Training procedure of MCRL}
    \label{alg1}
	\KwIn{ Dataset $\mathbf{X}$, training step $T$,mini-batch size $N$, the number of negative actions $M$, target network update rate $\sigma$,coefficients $\alpha$ and $\beta$,  temperature factor $\tau$, initialize V-networks $V_{\phi}$, policy extraction network $P_{\theta}$ with policy head $\pi$, reward function head $\hat{r}$ and state transition head $\hat{p}$,and target network $V_{\phi^\prime}$ with $\phi^\prime \leftarrow \phi$.}
	\KwOut{all parameters in the learning space $\phi,\theta$}
	Initialize all trainable parameters\; 
	\For{$t=1$ to $T$}{
	Sample mini-batch of $N$ transitions $(s,a,r,s')$ from $\mathbf{X}$\\
        Learn V function: $\phi \leftarrow \operatorname{argmin}_{\phi} L_V(\phi)$ (Eqn. \ref{Vloss})\\
        Randomly sample $M$ negative actions $\{a_i^-\}_{i=1}^{M}$\\
        Extract policy: $\theta \leftarrow \operatorname{argmin}_{\theta} L_P(\theta)$ (Eqn. \ref{Ploss})
	}
	
	return 
\end{algorithm}

\subsection{Dataset Description}
Here are the statistical features of the RetailRocket and Yoochoose datasets.
\begin{table}[t]
\caption{Statistics of datasets.}
\label{tab:dataStatics}
\renewcommand\arraystretch{1.1}
\begin{tabular}{cccc}
\hline
Dataset&RetailRocket&Yoochoose \\ \hline
\#sequences& 195,523&200,000\\
\#items&70,852&26,702 \\
\#clicks& 1,176,680&1,110,965\\
\#purchases& 57,269&43,946\\ \hline

\end{tabular}
\end{table}

\subsection{Figures of Ablation Study of Contrastive Learning and Hyperparameter }
Here are results of ablation study of  contrastive learning and hyperparameter.

\bibliographystyle{ACM-Reference-Format}
\bibliography{8_ref}


\end{document}